# OBSERVATION OF SINGLE TOP QUARK PRODUCTION AT THE TEVATRON

A.P. HEINSON

(for the DØ and CDF Collaborations)

*Department of Physics and Astronomy*
*University of California, Riverside, CA 92521-0413, USA*

This paper reports on the first observation of electroweak production of single top quarks by the DØ and CDF collaborations. At Fermilab's 1.96 TeV proton-antiproton collider, a few thousand events are selected from several inverse femtobarns of data that contain an isolated electron or muon and/or missing transverse energy, together with jets that originate from the decays of $b$ quarks. Using sophisticated multivariate analyses to separate signal from background, the DØ collaboration measures a cross section $\sigma\left(p\bar{p} \to tb + X,\, tqb + X\right) = 3.94 \pm 0.88$ pb (for a top quark mass of 170 GeV) and the CDF collaboration measures a value of $2.3^{+0.6}_{-0.5}$ pb (for a top quark mass of 175 GeV). These values are consistent with theoretical predictions at next-to-leading order precision. Both measurements have a significance of 5.0 standard deviations, meeting the benchmark to be considered unambiguous observation.

## 1  Single Top Quark Production Overview

At particle accelerators, top quarks are not produced very often because they are far heavier than all other known elementary particles. When they are created, it is most commonly in top-quark–top-antiquark pairs from the decay of a very energetic virtual gluon via the strong interaction. At a lesser rate (because the coupling is much weaker), top quarks and antiquarks are predicted [1] to be produced singly, without an antiparticle partner, from the decay of a highly energetic virtual $W$ boson via the electroweak interaction. There are three modes for single top quark production at a hadron collider [2]: t-channel production "*tqb*," where a $W$ boson and a $b$ quark fuse to produce the top quark, and there are a spectator light quark and a bottom antiquark; s-channel production "*tb*," where a $W$ boson decays to a top quark and bottom antiquark; and the production of a top quark together with a $W$ boson "*tW*," which occurs in both the s- and t-channels. The main t-channel process has a cross section of about 2 pb [3] (depending on the top quark mass value), the s-channel process's cross section is about 1 pb [3], and the *tW* process has a negligible rate at the Tevatron. The DØ collaboration announced evidence for the t-channel+s-channel processes combined at >3 standard deviation significance in 2006 [4] and this was also seen by the CDF collaboration in 2008 [5]. See Ref. 6 for details.

## 2 Event Selection

Since the first evidence publications, the DØ and CDF collaborations have extended the analyzed datasets and improved the analysis techniques. Signal-like events are selected with a high transverse momentum electron or muon, missing transverse energy, and between two and four jets. One or two of the jets must pass the $b$ identification criteria. The event yields after these selections are shown in the table in Fig. 1. CDF has an additional analysis channel with no identified charged lepton, which picks up events lost to lepton identification inefficiencies. The signal:background ratio is approximately 1:20 after all selections. The backgrounds are mostly from $W$+jets events (especially at low jet multiplicity), followed by $t\bar{t}$ pairs (especially at high jet multiplicity), with small contributions from $Z$+jets, dibosons, and multijets.

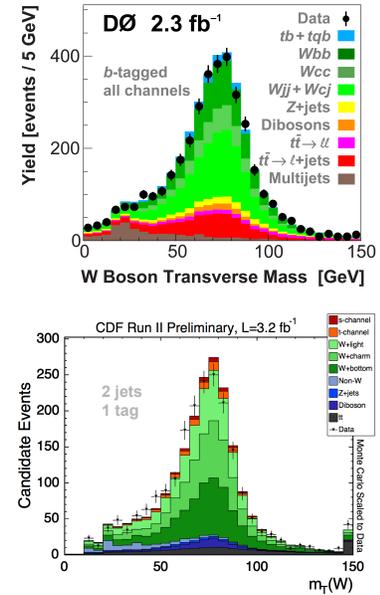

| Single Top Observation – Event Yields | | | |
|---|---|---|---|
| | **DØ** 2.3 fb$^{-1}$ | **CDF** 3.2 fb$^{-1}$ | **CDF** 2.1 fb$^{-1}$ |
| | Lepton+$\not{E}_T$+jets / $b$-tagged | | $\not{E}_T$+jets / $b$-tagged |
| $tb + tqb$ signal *1,*2 | 223 ± 30 | 191 ± 28 | 64 ± 10 |
| $W$+jets | 2,647 ± 241 | 2,204 ± 542 | 304 ± 116 *4 |
| $Z$+jets, dibosons | 340 ± 61 | 171 ± 15 | 171 ± 54 |
| $t\bar{t}$ pairs *1,*2, *3 | 1,142 ± 168 | 686 ± 99 | 185 ± 30 |
| Multijets | 300 ± 52 | 125 ± 50 | 679 ± 28 *5 |
| **Total prediction** | 4,652 ± 352 | 3,377 ± 505 | 1,403 |
| **Data** | 4,519 | 3,315 | 1,411 |

*1  DØ's $tb+tqb$ signal and $t\bar{t}$ background use $m_{top}$ = 170 GeV (and signal $\sigma_{(N)NLO}$)
*2  CDF's $tb+tqb$ signal and $t\bar{t}$ background use $m_{top}$ = 175 GeV (and signal $\sigma_{NLO}$)
*3  DØ's analysis includes 4-jet events, so the $t\bar{t}$ yield is higher
*4  CDF's $\not{E}_T$+jets channel $W$+jets yield does not include $Wjj$ where $j$ = light jet
*5  CDF's $\not{E}_T$+jets channel Multijets yield includes $Wjj$ events

Figure 1: The table shows the numbers of events after all selections have been applied. The plots show the $W$ boson transverse mass distributions for (a) DØ data with an electron or muon, two, three, or four jets, and one or two $b$ tags; (b) CDF data with an electron or muon, two jets, and one $b$ tag.

## 3 Background Model Checks

Extensive checks have been performed on the background models to ensure they reproduce the data in each analysis channel for all variables used in the multivariate analyses. To test the components of the background separately, data samples before $b$ tagging (dominated by $W$+light jets), with high jet multiplicity, $b$ tags, and large total transverse energy (dominated by top quark pairs), and with low jet multiplicity, only one $b$-tagged jet, and low total transverse energy (dominated by $W$+jets with half the jets from $b$ and $c$ decays), are examined in all variables for shape agreement between background model and data, and for the overall normalization. Good agreement is found, as shown in the example plots in Fig. 2.

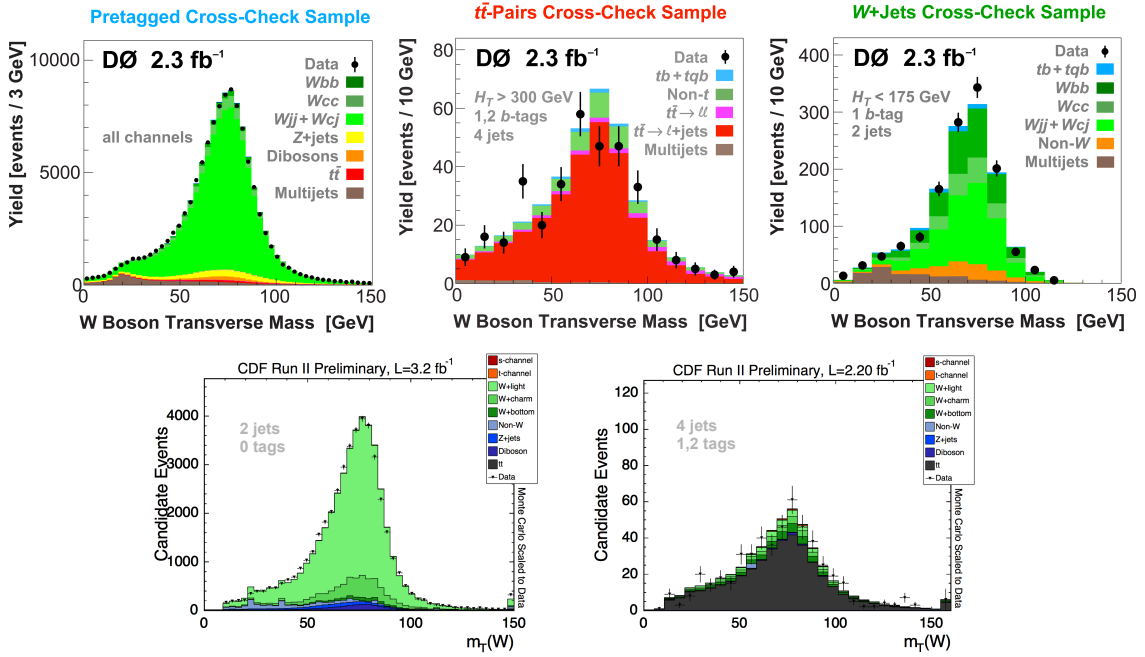

Figure 2: Upper row shows cross-check distributions for DØ's data. Lower row shows CDF's data.

## 4 Signal-Background Separation

Each collaboration uses several multivariate methods to separate the signal from background, since the signal forms too small a fraction of the data to just count events. These powerful uses of all available information about the events allow a significant signal to be isolated and the cross section measured with good precision. The discriminant output distributions are used in Bayesian binned likelihood calculations and then combined to make the final measurements. Figure 3 shows the final discriminant outputs from the combination of methods, for all analysis channels combined. The signal (light blue for DØ, salmon and tan for CDF) can be seen clustered near the high end of the distributions.

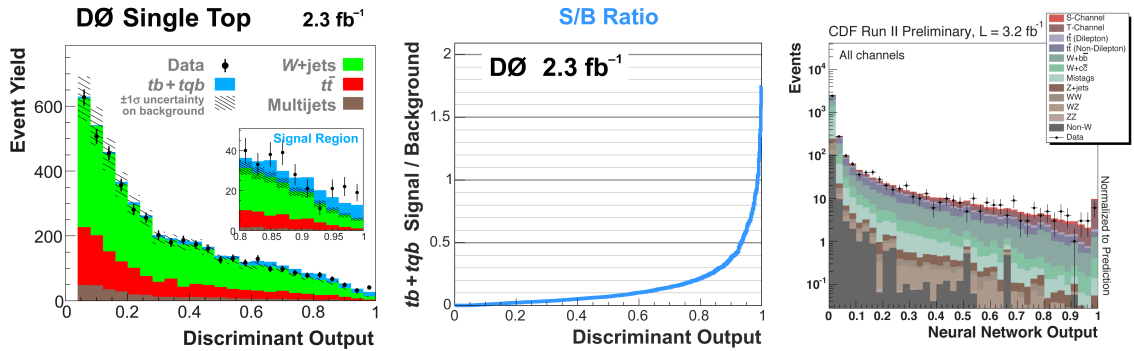

Figure 3: Left plot shows the final discriminant output distribution for DØ with all analysis channels combined. The center plot shows the signal:background ratio for this plot. The right plot shows the discriminant output distribution for all channels combined from CDF's analysis.

## 5  Results

The collaborations each combine their separate measurements using multivariate techniques. The results have been published [7,8] and are shown in Fig. 4. Recently, the measurements have been combined [9], and this result is also shown. The cross section measurements are used to extract limits on the value of the CKM matrix element $|V_{tb}|$ without assuming there are only three quark generations. With a significance of 5.0 standard deviations, these measurements form the first observation of electroweak production of single top quarks. For more details, see Ref. 10.

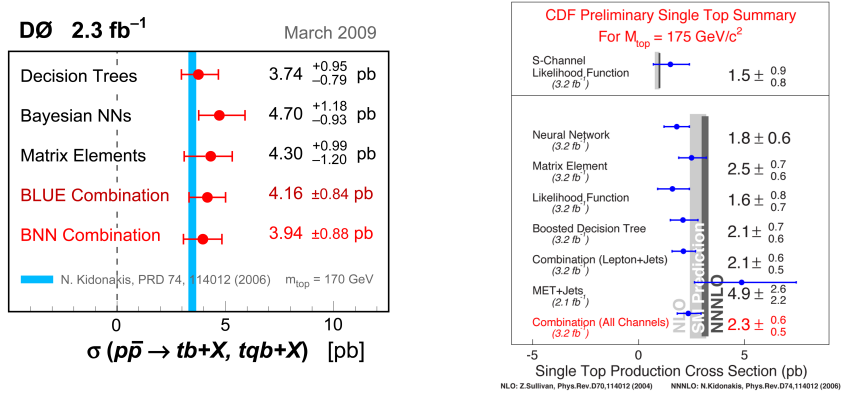

Figure 4: The plots show the separate single top cross section measurements, the table has the final results.

## References

1. S.S.D. Willenbrock and D.A. Dicus, Phys. Rev. D **34**, 15 (1986); C.-P. Yuan, Phys. Rev. D **41**, 42 (1990); S. Cortese and R. Petronzio, Phys. Lett. B **253**, 494 (1991).
2. A.P. Heinson, A.S. Belyaev, and E.E. Boos, Phys. Rev. D **56**, 3114 (1997).
3. N. Kidonakis, Phys. Rev. D **74**, 114012 (2006); B.W. Harris *et al.,* Phys. Rev. D **66**, 054024 (2002).
4. V.M. Abazov *et al.* (DØ Collaboration), Phys. Rev. Lett. **98**, 181802 (2007); V.M. Abazov *et al.* (DØ Collaboration), Phys. Rev. D **78**, 012005 (2008).
5. T. Aaltonen *et al.* (CDF Collaboration), Phys. Rev. Lett. **101**, 252001 (2008).
6. A.P. Heinson, in the proceedings of the 19th Hadron Collider Physics Symposium, arXiv:0809.0960.
7. V.M. Abazov *et al.* (DØ Collaboration), Phys. Rev. Lett. **103**, 092001 (2009).
8. T. Aaltonen *et al.* (CDF Collaboration), Phys. Rev. Lett. **103**, 092002 (2009).
9. The CDF and DØ Collaborations, arXiv:0908.2171.
10. http://www-d0.fnal.gov/Run2Physics/top/singletop_observation/Blois_Heinson_observation_talk.pdf
    http://www-d0.fnal.gov/Run2Physics/top/singletop_observation
    http://www-cdf.fnal.gov/physics/new/top/public_singletop.html